\documentclass[12pt]{article}

\usepackage{graphicx}

\titlepage
\begin{document}
\begin{center}
{ RESONANCE SPIN FILTER}\\
\vskip0.2cm {\large  B.S.\,Pavlov $^{1,2}$, A.M.\,Yafyasov $^1$}
\vskip5pt
\end{center}
\noindent $^1$ V.A. Fock Institute of Physics, St.Petersburg State
University, St.Petersburg, 198504, Russia\newline $^2$ Department
of Mathematics, University of Auckland, Private Bag 92019,
Auckland, New Zealand.
 \vskip0.2cm \centerline{\large\bf Abstract}

During last years the world-community of nano-electronics is
engaged in a  search of new physical principles, materials  and
technologies  on which  the  quantum  spin-transistor may be
manufactured \cite{DattaAPL}. This anticipated device  could
become a base  of the toolbox  of  quantum computations and help
testing the constructions of various quantum networks.
 \par
Basic  principle of the spin-transistor was suggested in the paper
\cite{DattaAPL}. The  leading  idea of  the proposal is  the use
of  the spin-orbital interaction \cite{Rashba} for  creation of
the spin-polarized  current  in the  transistor  channel.
Actually, to produce a real device based on the above mentioned
principle one should:

1. select  proper  material with maximal spin-orbital interaction
causing a considerable spin-orbital splitting,

2. suggest a method of introduction and withdrawal of  electrons
with certain spin polarization.

 \par An extended  analysis of  experimental and  theoretical ideas  leading to  the
solution of  the first  problem was  suggested  in the pioneering
paper \cite{Rad89}. According  to \cite{Rad96,RadYaf,RadYaf0} the
maximal spin  splitting  may  appear in  semiconductors  with Kane
dispersion for spectral bands. The  magnitude of splitting is
bigger for materials  with smaller  Kane  gap, see  \cite{Shur}.
Generically, the spin-orbital splitting  is two times  bigger in
the semiconductors  with  inverse band structure, compared with
other semiconductors, see \cite{RadYaf0,Adil1}.
  \par
In the  simplest phenomenological  Rashba  model for the
non-symmetric quantum well the corresponding additional linear
term with Rashba parameter $\alpha$ is  added  to  the quadratic
kinetic
 term:
  \begin{equation}
  \label{alpha}
E^{^{^{\pm}}} = \frac{\hbar^{2} k^{^2}}{2 m} \pm \alpha k.
\end{equation}
The  Rashba parameter   $\alpha$  is  defined  by  the  band structure of the
material, in particular by  the  magnitude  $\Delta_{_R}$ of the
gap, by the quasi-momentum, by  the electrostatic potential $V(z)$
forming  2D- electron gas and  by the shape of the wave-functions
of 2d electrons. According to  \cite{Rad89,Rad96,RadYaf}  both the
theory and the experiment vote in favor of  materials  Cd Hg Te,
where $\Delta_{_R}$ circa $40-60$ meV, and the  magnitude of the
effective Rashba  parameter is about $(0.2-0.3) 10^{^{-10}}$ ev/m.
This  is approximately $10$ times  bigger, that  the parameter
$\alpha$  in other hetero-structures studied  before :
$\Delta_{_R}\approx (0.02 : 5)$ meV and $\alpha \approx
(10^{^{-12}} - 10^{^{-{11}}})$ eV m. We guess that such  materials
as  Cd Hg Te are  most prospective for the high-temperature
Spintronics. Analysis done in \cite{RadYaf,RadYaf0,Adil1}  shows
that large values  of spin-orbital splitting are also achieved for
InAs and HgTe. All these semi-conductors are representatives of
the class of narrow-gap materials with the quasi-relativistic
dispersion function  and  wide spectral bands.
\par
In semiconductors with a nearly parabolic dispersion  curve  $E
\approx \frac{\hbar^2 p^2}{2 m*}$ the magnitude of the expected
splitting is at least 2 or 3 orders less  than in  narrow-gap
materials  listed  above, hence it is hardly accessible for
observations. Nevertheless in actual note we suggest a theoretical
analysis of resonance transmission across the quantum well in a
material with parabolic dispersion curve  and  the  Rashba
Hamiltonian just included additively as in (\ref{alpha}), having
in  mind  that it is universal  for  low  temperature  and small
$\alpha$. We reveal effects caused by the shape of the oscillatory
modes in resonance processes, thus making  a step toward the
solution of the second of above problems - the problem of
introducing and withdrawal of spin-polarized electrons, - actually
the problem of registration of  the spin-polarization.
  We calculate transmission coefficients for the  Resonance Spin Filter
designed in form of a quantum network consisting of a quantum well
with three semi-infinite quantum wires (an input  wire  and  two
terminals) attached  to it. Transmission of electrons across the
well from the input quantum  wire to  terminals is caused by the
excitation of the resonance  oscillatory mode in the well. The
resonance mode is non-symmetric with respect to the spin-inversion
and hence  the  spin-selection can be  achieved via special choice
of the geometry of the  well and  contact points of terminals on
the boundary of it.
   \par
Analysis of the spin-independent resonance transmission was done
in \cite{MP00,boston}, where optimization of transport properties
was achieved based  on  distribution of zeroes of  the  normal
derivatives  of the  resonance  mode on the border of the  well.
An essential difference of the actual problem from the previous
one is  the fact, that the resonance eigenfunctions in actual
problem are {\it complex} two-component spinors, hence the
corresponding  one-pole approximation for the Scattering matrix,
see (\ref{onepole}) below, is presented via $2 \times 2$ block's
which contain characteristics of the shape of the  resonance mode
on the  border of the  well and  play the roles of
transmission-reflection coefficients for electrons with spin up or
down. Nevertheless the optimization of selection can be achieved
based on the explicit formula (\ref{onepole}).
\par
\section{Hamiltonian and  the Intermediate Hamiltonian}
Consider a  network $\Omega=\Omega_0 \cup \omega_1\cup\dots$
 on  $(x,z)$  plane  combined  of  a circular quantum well
 $\Omega_0$ and three straight semi-infinite quantum
wires $\omega_s,\,\, s= 1,2,3$ of  constant  width  $\delta$
attached  to the well $\Omega_0$  such that  the orthogonal bottom
sections $\gamma_{_{s}}$ of  the  wires $\omega_s$ are parts of
the piece-wise smooth boundary  $\partial \Omega_{0}$ of  the well
$\Omega_{0}$ . We consider scattering of  electrons in the network
in presence of  a  strong  electric  field  directed orthogonally
to  the $(x,z)$ plane. The  wave-function $\Psi$ of the  electron
is  presented  by  spinor $(\psi_1,\,\,\psi_2)$, and the
spin-orbital  interaction is  taken in  form of Rashba Hamiltonian
\cite{Rashba} : a  cross product - of  the  vector $\sigma$ of
Pauli matrices and the vector $p$ of  momentum of  electron :
\begin{equation}
\label{spinorb} H_{_R}= \alpha\left[\sigma,\,\, p \right],
\end{equation}
where $\alpha$ is  an absolute  constant. In presence of  a strong
electric field ${\cal E} = |{\cal E}| e_{_{y}} $ directed
orthogonally to the  $(x,z)$ plane the corresponding
Schr\"{o}dinger equation on the well is
\begin{equation}
\label{Schredinger} L u = -\frac{\hbar^2}{2 m*}\bigtriangleup u +
V u  +
 \alpha[\sigma_z \ p_x - \sigma_x \ p_z] u
\end{equation}
where  $\alpha[\sigma_z \ p_x - \sigma_x \ p_z] = \left(
H_{_R}\right)_{_y}$ is the $y$-component of  the  cross product
$\alpha\left[\sigma,\,\, p \right]$, and the linear potential $ V
: = |{\cal E}| y $  on the  well is defined by the macroscopic
electric field ${\cal E}$. The above Rashba Hamiltonian is
non-self-adjoint in the space $L_2 (\Omega)$ but the  whole
Schr\"{o}dinger operator (\ref{Schredinger}), is  a self-adjoint
operator in $L_2 (\Omega)$ on the domain of sufficiently smooth 
functions  with appropriate  conditions  imposed 
on   their  boundary  values, see  (\ref{boundcondG},\ref{boundcondR}). On the
wires the potential is constant $V = V_{\infty}$ , but we assume
that the Schr\"{o}dinger equation on the wires $\omega_{_{s}} =
\left\{  0< \eta_{_{s}}< \delta,\, 0< \xi_{_{s}} <\infty\right\}$
contains an anisotropic tensor of effective mass:
\begin{equation}
\label{Schrwires} l u = -\frac{\hbar^2}{2
m^{\parallel}}\,\frac{d^2 u}{d \xi ^2} - \frac{\hbar^2}{2
m^{\bot}}\,\frac{d^2 u}{d \eta ^2} + V_{\infty} u ,
\end{equation}
and the  width of wires is constant and equal to $\delta$. We
neglect the  spin-orbital interaction in the  wires. On the sum
$\Gamma = \sum_{s=1}^3 \gamma_s$ of bottom sections of the wires,
separating the wires from the  well, we  impose proper matching
boundary conditions on the  boundary of the  well:
\begin{equation}
\label{boundcondG} 
\frac{\hbar^2}{2 m^*}\,\,\,\frac{\partial
u}{\partial n} - \frac{i \alpha}{2} \left[ \sigma,\, n\right]u
\bigg|_{\partial \Omega_{_{0}} \backslash \Gamma} = 0,
\end{equation}
\begin{equation}
\label{boundcondR}
 \frac{\hbar^2}{2 m^*}\,\,
\frac{\partial
u}{\partial n} - \frac{\hbar^2}{2
m^{\parallel}}\,\,\,\frac{\partial u_s}{\partial n} - \frac{i
\alpha}{2} \left[ \sigma,\, n\right]u \bigg|_{\gamma_s} = 0,
\end{equation}
They  define  a  self-adjoint operator  ${\cal L}$ on  $L_2
(\Omega)$  which plays  a  role  of the  Hamiltonian  of  the
electron on the  network. Following the pattern of \cite{boston}
we consider the scattering problem for ${\cal L}$  on the network
$\Omega$ and calculate the transmission coefficients from the
input wire $\omega_1$ to the terminals $\omega_2,\,\, \omega_3$
across the well and  estimate the quantum conductance on resonance
energy, based on Landauer formula,see \cite{Landauer70, Buttiker85}.
\par
Denote by $e_l = \sqrt{\frac{2}{\sqrt{2 m^{^{\bot}}}\delta}}
\sin \frac{\pi l \eta}{\sqrt{2 m^{^{\bot}}}\delta},\,\,
l=1,2,3,\,\,\dots,\, 0<\eta<\delta$  the eigenfunctions of  the
cross-sections  of  the  wires and assume  that  the  Fermi level
in the  wires  lies in the  middle of the  first  spectral band in
the  wires  $E_{_F} = V_{\infty} + \frac{5}{2} \frac{\hbar^2}{2
m^{\bot}}\frac{\pi^2}{\delta^2}$. Denote  by  $P_+ $ the
orthogonal  projection onto the linear  hull  of  the entrance
vectors $\bigvee_{_{s}} e_{_{1,s}} =  E_{_{+}} $ of the open first
channel, and by  $P_{-}$ the complementary projection in $L_2
(\Gamma) \,\, : \,\,I = P_{+} + P_{-}$. We define the {\it
Intermediate  Hamiltonian}  $\hat{\cal L}$ by the same
Schr\"{o}dinger differential expressions (\ref{Schredinger},
\ref{Schrwires}) but replacing the matching conditions
(\ref{boundcondR})  by  the  ``chopping-off''  boundary conditions  in open  channels:
\begin{equation}
\label{match+} P_+  u_s\big|_{_{\gamma_s}} = 0,\, \,s=1,2,3.
\end{equation}
and the  matching conditions in closed  channels :
\begin{equation}
\label{match-}
P_- [u_0 - u_s]\,\,\big|_{_{\gamma_s}} = 0,\,\,
 \frac{\hbar^2}{2 m^*}\,\,\,P_-\frac{\partial
u_0}{\partial n} - \frac{\hbar^2}{2
m^{\parallel}}\,\,\,P_-\frac{\partial u_s}{\partial n} - \frac{i
\alpha}{2} \left[ \sigma,\, n\right] P_- u_0 \bigg|_{\gamma_s} =
0.
\end{equation}
 Note that the boundary term  $\frac{i \alpha}{2} \left[
\sigma,\, n\right]$ arising from the Rashba Hamiltonian commutes
with projection onto the entrance vectors  of  the channels. The
operator $ {\hat{\cal L}}$ on the network defined by  the above
differential expressions (\ref{Schredinger}, \ref{Schrwires}) with
the boundary conditions (\ref{match-},\ref{match+}) and the
Meixner conditions at the inner angles of the domain  is
self-adjoint. The  operator $\hat{\cal L}$ is split  as an
orthogonal sum  $\hat{\cal L} = {\bf l}_{_{1}} \oplus L_{_{R}}$ of
the operator ${\bf l}_{_{1}}$  on the open channel in the wires and the operator
$L_{_{R}}$ acting in the orthogonal complement  of  the  open channels. 
This  operator  plays  a  role  of an intermediate  operator .  We  will  derive  
an explicit  formula  for  the  scattering matrix   in terms  of    spectral data  
of  $L_{_R}$, see  (\ref{Smatrix}). The spectrum of the part
${\bf l}_{_1}$ of $\hat{\cal L}_{_{R}}$ in the  open  channels  is
just a semi-axis $V_{\infty}+ \frac{\pi^2
\hbar^{^2}}{2m^{\bot}\delta^2} < \lambda < \infty.$ The
absolutely-continuous  spectrum of  the part ${L}_{_{R}}$ of
the operator ${ L}_{_{R}} = \hat{\cal L}\ominus {\bf
l}_{_1}$ on the  orthogonal complement of the  open  channels 
consists  of   a  countable  family of branches
$\cup_{l=2}^{\infty}\left[\left.\frac{l^2 \hbar^{^2} \pi^2}{2
\mu^{\bot}\delta^2} +
 V_{\infty},\,\infty \right.\right)$.
\vskip0.5cm
\section{Scattering  matrix}
Denote by  $G_{_{R}} $ the  Green function of the   the  operator
${L}_{_{R}}$. The  solution   ${\bf u}$  of  the Dirichlet problem for the
former equation with  the  data $\left\{ u_{_{1}},\, u_{_{2}},\, u_{_{2}}
\right\}: = u_{_{\gamma}}\in E_+$  on  $\Gamma$  can be  presented  as
 \[
u (x)= - \int_{\Gamma}\left( \frac{\hbar^2}{2 \mu^*}\frac{\partial
G_{_R} (x,s)}{\partial n_{_{s}}} - i\alpha [\sigma,
n_{_{s}}]G_{_R} (x,s) \right) u_{_{\gamma}} (s) d\Gamma := {\cal
P}  u_{_{\gamma}}.
 \]
We match  $u$  with  the
Scattering  Ansatz
 ${\bf u} = e^{-i p \xi} \nu + e^{i p \xi} S \nu,\, \nu \in E_+,\,\,
p = \frac{\sqrt{2 \mu^{^{\parallel}}}}{\hbar} \,\,\sqrt{\lambda -
V_{_{\infty}} - \frac{\pi^{^{2}}}{2 \mu^{^{\bot}} \,
\delta^{^{2}}}}$  in the  first  channel :
\[
 \frac{\hbar^2}{2 m^*}
\,\,\,P_+\frac{\partial u}{\partial n} -
\frac{i
\alpha}{2}
 \left[ \sigma,\, n\right] P_+ u \bigg|_{\gamma_s} =
 \frac{\hbar^2}{2 m^{\bot}}\,\,\,P_+\frac{\partial {u}_{_{s}}}{\partial
n}.
\]
Taking  into  account  the  continuity  $[u-{ u}_{_{s}}]
\bigg|_{_{\gamma_{_{s}}}} = 0$ and  denoting by  ${\cal D}_{_R}$
the boundary  differential operation  $ {\cal D}^{^R}_{_{x}} u =
\left( \frac{\hbar^2}{2 \mu^*}\frac{\partial u }{\partial n} -
i\alpha [\sigma, n] u \right) $ and by  $\Lambda_R (\eta,\eta')$
the generalized kernel  of  the corresponding Dirichlet-to-Neumann
map (DN-map) $\Lambda_R $, see \cite{SU2}, of the  operator
${L}_{_R} $ on $\Gamma$ : $ \Lambda_{_R} (\eta, \eta')=  -
{\cal D}^{^R}_{_{x}}\,\, {\cal D}^{^R}_{_{x'}} \,\, G_{_R}
(x,\,x')$, with $x|_{_{\Gamma}} = \eta,\, x'|_{_{\Gamma}} =
\eta'$, we   calculate the DN map $P_{_{1}}\Lambda_R P_{_{1}}: =
\Lambda_R^{^1}$ framed by  the orthogonal projections onto  the
entrance vectors $e_{_{1,s}}$ of the  first  channel :
  $P_{_{1}} = \sum_{_{s = 1}}^{^3} e_{_{1,s}}\rangle \,\,\langle e_{_{1,s}}$.
This  gives the  following  formula   for  the  scattering  matrix
:
\begin{equation}
\label{Smatrix} S = - \frac{\Lambda_R^{^1}  + ip I}{\Lambda_R^{^1}
- ip I}.
\end{equation}
The   transport   properties   of the  filter  for given temperature $T$ are 
 defined   via  averaging of the   corresponding  transmission coefficients
 $S_{_{s1}}$ over  Fermi  distribution   on the  essential spectral  interval, 
 $\Delta_{_{T}} = \left\{E_{_{F}} - \kappa T <\lambda< E_{_{F}} +
\kappa T\right\} $ , see for  instance  \cite{Aver_Xu01}.
We  may  obtain  a  reasonably  good  approximation for  the
Scattering  matrix, substituting for $\Lambda_{_R} (x,y)$  the
corresponding  spectral  sum over  all  eigenvalues $\lambda_m$ of
the operator $\hat{L}_{_R}$  on the  essential spectral  interval, if 
the interval 
does not overlap with the continuous spectrum of $L_{_{R}}$:
\begin{equation}
\label{essential}
 \Lambda_{_R}^{^1} (\eta, \eta') \approx
-\sum_{\Delta_{_{T}}} \,\,\frac{P_{_{1}}{\cal D}^{^R}_{x}\,\,
\varphi_m (\eta)\rangle \,\,\langle \,
 P_{_{1}}{\cal D}^{^R}_{x'}\,\, \varphi_m (\eta')}{\lambda_{_m} - \lambda
},\,\,\,\,\, \eta, \eta' \in \Gamma.
\end{equation}
The formulae  (\ref{Smatrix},\ref{essential}) show that the
eigenvalues and eigenfunctions of the intermediate operator define
the structure of  the Scattering Matrix on $\Delta_{_{T}}$ and  the  transport
 properties  of the  spin-filter  based  on the  quantum well.  Recovering  
 necessary   spectral  data  of the  intermediate  operator   with   the 
 non-standard   boundary  conditions
 (\ref{match+},\ref{match-})  cannot  be  done  with existing   commercial  software 
 and   needs   creation of  special programs. 
 \par 
  Assume
that there exist a resonance eigenvalue $\lambda_{_0}$ of the
operator $\hat{L}_{_R}$ which is equal to the  Fermi -level in the
wires $\lambda_{_0} = E_{_F}$. Then on a (small) part of the
essential spectral interval defined by  the temperature we may
substitute the DN-map by  the resonance term  only.:
  \[
\Lambda_{_R}^{^1} (\eta,\eta') \approx \Lambda_{_{essential}}^{^1}
(\eta,\eta') - \frac{P_{_{+}}{\cal D}^{^R}_{_x}\,\, \varphi_0
(\eta)\rangle \,\,\langle \,
 P_{_{+}}{\cal D}^{^R}_{_{x'}}\,\, \varphi_0 (\eta')}{\lambda_{_0} - \lambda }
\]
The  residue of the resonance polar term is  proportional to the
projection  onto the  one-dimensional subspace  spanned  by the
portion
 $P_{_{+}}{\cal D}^{^R}_{{x}}\,\, \varphi_0 (\eta)\,\bigg|_{_{\gamma_s}} = \phi_{_{s}}$
of the  resonance eigenfunction  in the entrance  subspace of  the
wire $\omega_{_{s}}$:
\[
\Lambda^{^{1}}_{_{R}} (\eta,\, \eta') \approx
\frac{P_{_{1}}{\cal D}^{^R}_{_{x}}\,\, \varphi_0 (\eta)\rangle \,\,\langle
\,
 P_{_{1}}{\cal D}^{^R}_{_{x'}}\,\, \varphi_0 (\eta)}{\lambda - \lambda_{_0}}
  = |{\bf \phi}|^{^{2}} \left\{
P_{_{\phi}}\right\}_{_{s,\, s'}},
\]
where   $\phi = \oplus\sum_{_{s=1}}^3 \, \phi_{_{s}}$ , 
  $|\phi|^{^{2}} =  \sum_{_{s=1}}^{^{3}} |\phi_{_{s}}|^{^{2}}$.
  This  gives  the  
one-pole   approximation  for the scattering matrix :
  \begin{equation}
 \label{onepole}
S (\lambda) \approx S_{_{approx}} (\lambda) = P^{^{\bot}}_{_{\phi}} -
\frac{|\phi|^{^2} + i p (\lambda - \lambda_{_0})} {|\phi|^{^2} - i
p (\lambda - \lambda_{_0})} \,\, P_{_{\phi}},
 \end{equation}
where $P^{^{\bot}}_{_{\phi}} = I - P_{_{\phi}} $. The  last
formula implies the  following  expressions  for transmission
coefficients at the  resonance energy $\lambda_0 =  E_{_{F}}$
  for  low  temperature:
\[
S_{_{s1}} (\lambda_0) \approx ( S_{_{approx}})_{_{s1}} (\lambda_0)= - 2
\frac{{\bf \phi}_{_{s}}\rangle \, \langle
\,{\bf \phi}_{_{1}}} {|{\bf \phi }|^{^2}}.
\]
Hence the  transmission coefficient from the  input  wire
$\omega_{_1}$ to the wire $\omega_{_s}$ is  represented  by  the
$2\times 2$ matrix constructed  as  a product of  spinors, and the
spin-dependent current   is  calculated based on the relevant
version of the Landauer formula, see \cite{Buttiker85}.  In   \cite{MP01,boston}
  the  resonance eigenfunctions  are  scalar and  real.
  Then the  transmission coefficients  are  defined  by 
  the  integrals 
  \[
 \int_{_{\gamma_{_{s}}}}  e_{_{\eta}}\, \frac{\partial \varphi_{_{0}} }{\partial n}(\eta) d\eta\,  \int_{_{\gamma_{_{1}}}}  e_{_{\eta}}\,  \frac{\partial \varphi_{_{0}} }{\partial n}(\eta) d\eta.
  \]
   The  selectivity  of the  devices  in    \cite{MP01,boston} is  guaranteed   by  the  
  presence of the  zeros  of  $ \frac{\partial \varphi_{_{0}} }{\partial n}$ in  
  the  middle  of the  bottom section  $\gamma_{_{s}}$ of the   wire   $s$.
  In  our  case   the  intergrands   in   the  corresponding  integrals  are  spinors,
  and  the  transmission coefficients  are   presented  by  matrices .   Using the
   superscripts   $\pm $ for  components of  spinors   with  spin  $\pm \, 1/2$
   we   write  down   the   formulae   for  transmission coefficients , for  instance:
 \begin{equation}  
 -\frac{1}{2} |\phi|^{^2} S^{^{++}}_{_{s1}} (\lambda_{_{0}}) =
  \int_{_{\gamma_{_{s}}}}  e_{_{1}} (\eta)\, \left( {\cal D}_{x}^{^R}\,\, \varphi_0  \right)^{^{+}}\bigg|_{_{\gamma_{_{s}}}}
  (\eta) d\eta\, 
   \int_{_{\gamma_{_{1}}}}  e_{_{\eta}}\,  \left( {\cal D}_{x}^{^R}\,\, \varphi_0 \right)^{^{+}}\bigg|_{_{\gamma_{_{1}}}}(\eta) d\eta.
 \end{equation}
The  magnitude  of the  transmission coefficients  defines  the  selectivity 
of the  spin-filter.  But    recovering  of  details  of the  parameter  regimes  of the  switch  
now  is   more  complicated  task,  that  in   \cite{MP01,boston} , since  
the  matrices  are  complex   and   the  selectivity  is  not  defined  by  zeros  
of  one  real function.  Nevertheless,  if the  geometry of  the  well and  the
positions of the  contact  zones  $\gamma_{_s}$ are chosen such
that, for  given electric  field ,  the   matrix  elements  $S^{^{++}}_{_{s1}} (\lambda_{_{0}}),\,\,S^{^{+-}}_{_{s1}} (\lambda_{_{0}})$   bigger  than
$S^{^{-+}}_{_{s1}} (\lambda_{_{0}})\,\,\,S^{^{--}}_{_{s1}} (\lambda_{_{0}})$, then
the  electrons  with   the  spin up  prevail  in the  exit   wire  $s$, for   non-polarized  
incoming  flow  in  the  wire $1$. 
 Based on the above formula (\ref{onepole}) one
can  calculate also the position of  the  resonance in the complex
plane ( the pole of the  Scattering Matrix), which   essentially  defines  the  speed  of  switching. 
 Based  on  the  one-pole  approximation  one  can  construct   a solvable  model  , \cite{Kurasov}
 of the  spin filter,   in form of a    quantum graph with   the  resonance   boundary 
 conditions  at  the  node. 
   \end{document}